\documentstyle[aps,epsf,twocolumn,prl]{revtex} 
\begin{document} 
\title{
Interacting Dark Resonances: Interference Effects Induced by Coherently 
Altered Quantum Superpositions} 
\author{M.~D.~Lukin$^{1,3,4}$, S.~F.~Yelin$^{1,2,3}$, M.~Fleischhauer$^{1,2}$,
and M.~O.~Scully$^{1,3}$} 
\address{$^1$ Department of Physics, Texas
A\&M University, College Station, TX 77843}
\address{$^2$ Sektion Physik, Ludwig-Maximilians-Universit\"at M\"unchen,
Theresienstra\ss e 37, D-80333 M\"unchen, Germany}
\address{$^3$ Max-Planck-Institut f\"ur Quantenoptik, 
85748 Garching, Germany}
\address{$^4$ ITAMP, Harvard-Smithsonian Center for Astrophysics, 
60 Garden Street, Cambridge, MA 02138}
\maketitle 
\begin{abstract} 	
We predict the possibility of sharp, high-contrast resonances in 
the optical response of a broad class of systems, wherein interference effects 
are generated by coherent perturbation or interaction of  
dark states. 
The properties of these resonances can be manipulated to design 
a desired atomic response. 
\end{abstract} 
The phenomenon of ``dark resonances'' or coherent population trapping
\cite{ari} is by now a well
known concept in optics and laser spectroscopy. It is 
a basis for such 
effects  as electromagnetically induced transparency 
(EIT) \cite{eit} and 
its applications to nonlinear optics \cite{nlo}, lasing without inversion 
\cite{lwi},
the resonant enhancement of the refractive index \cite{hi,indexp},
adiabatic population transfer \cite{adi}, sub-recoil laser cooling
\cite{ct} and atom interferometry \cite{chu}.

The essential feature of dark resonances is the existence 
of quantum superposition states which are decoupled from the coherent 
and dissipative interactions. As a general rule, interactions
involving such a ``dark state'' lead to decoherence and are undesirable.
In the present paper we demonstrate a new and qualitatively 
different approach to these problems. The present approach involves
multiple quantum superposition states that are coupled and interact 
{\it coherently}. We find that such interacting superpositions can be 
used, in many instances, to mitigate various decoherence effects and
to enlarge the domain of dark-state based physics.  

In particular, we show that coherent interaction leads to a splitting of 
dark states and 
the emergence of novel, sharp spectral features. 
While separate parts of the resulting optical response can be explained 
in terms 
of different, appropriately chosen superposition basis, their 
simultaneous presence and 
hence the ``double-dark'' resonance  structure as a whole 
is a definite 
signature of a new type of quantum interference effect.  The
phenomenon of interfering  
double-dark 
states is very general and occurs in a broad class of 
multi-state systems. The
effect can be induced, for 
example,  by a microwave field driving a magnetic dipole
transition, by 
optical fields inducing multiple two-photon transitions, by a static field,
or by a non-adiabatic coupling mechanism in
time-dependent laser fields \cite{Mik-Ar}. 

We show that the new resonances associated with 
the double-dark states can be made absorptive or transparent and their 
optical properties such as width and position can be 
manipulated by adjusting the coherent interaction. Furthermore,
a very weak incoherent excitation of the atoms can be sufficient to turn 
the absorption into optical gain. We anticipate that such ``designed''
atomic response can be of interest in areas as diverse as 
enhancement of optical activity in dense media, high resolution
spectroscopy, quantum well lasers,
and quantum computation. 

\begin{figure}[htb]
\epsfysize=5.5cm
\center{\leavevmode\epsfbox{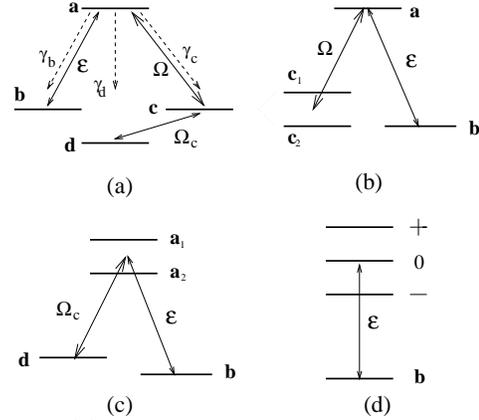}}
\caption{ (a) Four state system displaying double-dark resonances.
Unitary equivalent systems: (b) corresponds to generic model Fig.~1a 
after diagonalizing
the interaction with the coherent perturbation $\Omega_c$ (states
$\left|c_1\right\rangle$ and $\left|c_2\right\rangle$ correspond to
$(\left|c\right\rangle\pm\left|d\right\rangle)/\sqrt{2}$, 
respectively) and (c) after diagonalizing the 
interaction with the drive field $\Omega$ (states $\left|a_1\right\rangle$ and
$\left|a_2\right\rangle$ correspond to a pair
$(\left|a\right\rangle\pm\left|c\right\rangle)/\sqrt{2}$ 
displaying Fano interference due to spontaneous emission) (d) Dressed
state picture corresponding to the Figs.~a,b,c.}
\end{figure}

The basic physical mechanism leading to the novel response 
can be understood by considering the generic four-state system of Fig.1a. 
Here a resonant driving field and a weak probe field with
Rabi-frequencies $\Omega$ and ${\,{\cal E}}$  
couple two lower metastable states $c$ and $b$ with upper level $a$
and therefore form a simple $\Lambda$ configuration. The resulting
dark state is coherently coupled by a real or effective coherent 
field with Rabi-frequency $\Omega_c$ to another metastable state 
$d$. As noted above a variety of different mechanisms can cause this coupling. 
This model is quite  general since it is unitary equivalent to a broad class
of other 4-state systems, some of which are shown in Fig.1. All of these 
schemes are described by the identical semiclassical dressed-state 
picture (Fig.1d) , which provides useful insight into the origin of 
the interference between double-dark resonances.  

Let us begin with the system of Fig.1a, in which all 
coherent processes are described - within the rotating 
wave approximation -  by the following Hamiltonian matrix 
\begin{equation}
{\cal H}_0= 
-\hbar\pmatrix{\Delta_0 &  \Omega  &          0 &        {{\cal E}} \cr
               \Omega &        0  &  \Omega_c &           0 \cr
                    0 & \Omega_c & -\Delta_c &           0 \cr
                  {{\cal E}} &        0  &          0 & \Delta_0-\Delta\cr},
\end{equation}
corresponding to the vector of state amplitudes 
$(c_a,c_c,c_d,c_b)$, ($c_\mu:=\left\langle\mu\right|\psi\rangle$). Here
$\Delta = \nu_p-\omega_{ab}$, $\Delta_0 = \nu-\omega_{ac}$, and 
$\Delta_c = \nu_c - \omega_{cd}$ are the detunings of the probe field,
the drive field, and the coherent perturbation.  The three dressed sublevels, 
generated by the driving fields $\Omega$ and $\Omega_c$, 
and the corresponding frequencies
read to first  order in $\Omega_c$
\begin{eqnarray}
\left|+\right\rangle &=& \frac{1}{\Omega_0}
\left[-\omega_+\left|a\right\rangle
+\Omega\left(\left|c\right\rangle+{\Omega_c \over \omega_0
-\omega_+}\left|d\right\rangle\right)\right],\\ 
\left|-\right\rangle &=&  \frac{1}{\Omega_0}\left[\Omega
\left|a\right\rangle+
\omega_+
\left(\left|c\right\rangle+{\Omega_c \over \omega_0
-\omega_-}\left|d\right\rangle\right) 
\right],\\
\left|0\right\rangle &= & \left|d\right\rangle -
\frac{\Omega_c\Omega}{\widetilde\Omega^2}\left|a\right\rangle 
           +\frac{\Omega_c(\Delta_c+\Delta_0)}{\widetilde\Omega^2}
\left|c\right\rangle, 
\end{eqnarray}
\begin{eqnarray}
\omega_\pm &= & -\frac{\Delta_0}{2} \mp\sqrt{\Omega^2
+\frac{\Delta_0^2}{4}}, \\ 
\omega_0 &= & \Delta_c,
\end{eqnarray}
where $\Omega_0^2:=\Omega^2 +(\Delta_0/2 +\sqrt{\Omega^2+\Delta_0^2/4})^2$, 
and we have assumed a sufficiently large splitting between the dressed
state energies such that $|\widetilde\Omega^2|:=|\Omega^2 -
\Delta_c(\Delta_c+\Delta_0)|\gg \Omega_c^2$. 
Two of the dressed states, $\left|\pm\right\rangle$, correspond, in the
limit of vanishing perturbation $\Omega_c \rightarrow 0$, to the usual
Autler-Townes dressed components split by $2\sqrt{\Omega^2+\Delta_0^2/4}$.
Since both have a finite overlap with the excited state $\left|a\right\rangle$,
there is quantum interference in the absorption or spontaneous
emission on the probe-transition leading to a single dark resonance. 
The third dressed state $\left|0\right\rangle$
coincides in this limit with the bare state $\left|d\right\rangle$ and
hence is  
decoupled from the system. This is no longer so in the presence of a 
second drive field $\Omega_c$. 
In this case the dressed state $\left|0\right\rangle$ contains an admixture of 
$\left|a\right\rangle$ and thus has a non-zero dipole matrix 
element with ground state $\left|b\right\rangle$. From this coupling
result transitions 
between $\left|b\right\rangle$ and $\left|0\right\rangle$ corresponding to
three-photon hyper-Raman resonances as well as additional 
interference effects. The latter
lead to a new pair of
transparency points, as shown in Fig.2a. 

Alternatively, the system can be analyzed by diagonalizing 
the interaction with the coherent coupling ($\Omega_c$), which leads to the 
system shown in Fig.1b. It corresponds to a four-level system with two 
drive fields of Rabi-frequency $\Omega/\sqrt{2}$ forming two
different $\Lambda$ subsystems with lower states
$\left|c_{1,2}\right\rangle$ split 
by $\pm\Omega_c$. 
In such a system, the existence of two distinct dark resonances is clear
at hand, each corresponding to a two-photon resonance between
$\left|b\right\rangle$ and states $\left|c_1\right\rangle$ and 
$\left|c_2\right\rangle$ respectively. 
In this basis the central narrow structure is due to 
interference induced by the coherent interaction between the two dark states. 
\begin{figure}
\epsfysize=2.5cm
\center{\leavevmode\epsfbox{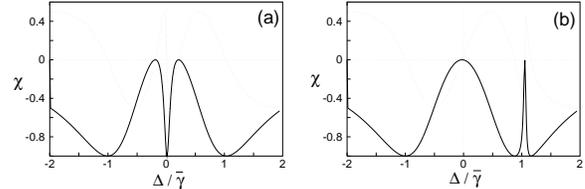}}
\caption{Imaginary (solid lines) and real (dashed lines) parts of the 
probe susceptibility in the units of $\eta/\bar{\gamma}$ for a system 
of Fig.~1a. Parameters are $\gamma_b=\gamma_c=\gamma_d=\bar{\gamma}$, 
$\Omega = \bar{\gamma}$, $\Omega_c = 0.2\bar{\gamma}$,
$r=\gamma_0 = 0$, $\Delta_0 = 0$. (a) $\Delta_c=0$. (b) $\Delta_c =
\bar{\gamma}$. Case  (b) is the special case when the dressed energies 
intersect ($E(\left|-\right\rangle)\sim E(\left|0\right\rangle)$).}
\end{figure}

Hence, in either basis quantum interference is an essential feature of 
interacting dark resonances.  This quantum interference combined with the 
possibility of tuning the position
of the state $\left|0\right\rangle$ and adjusting its coupling
strength  allows one 
to ``engineer'' the 
atomic response. We now discuss this in more 
detail.

To quantify the properties of double-dark resonances we examine the response 
of the system (Fig.1a) using the full set of density matrix equations.
We assume a weak probe field, and begin with 
case when all atoms are in ground state $b$ .  We also take into account
transit-time broadening with a corresponding rate $\gamma_0$.
The linear susceptibility is then given by:
\begin{eqnarray}
\chi = 
\frac{{\rm i}\eta \; \; \Gamma_{\!cb}}{\Gamma_{\!ab}\Gamma_{\!cb}
+\Omega^2} \biggl( 1 + 
\frac{\Omega^2}{\Gamma_{\!cb}}  
\frac{\Omega_c^2}{\Gamma_{\!ab}(\Gamma_{\!cb}\Gamma_{\!db}  +
\Omega_c^2) + \Omega^2 
\Gamma_{\!db}}
\biggr),
\label{lin1}
\end{eqnarray}
where  $\eta =\gamma_b
3{\cal N} \lambda^3/(8\pi^2)$,
${\cal N}$ is the
atomic density and $\wp$ is the dipole matrix element of the probe transition. 
$\gamma_{ij}$ are the relaxation rates of
the respective  coherences, and $\Gamma_{\!ab} = \gamma_{ab} + {\rm
i}\Delta$, $\Gamma_{\!cb} = 
\gamma_{cb} + {\rm i}(\Delta - \Delta_0)$, $\Gamma_{\!db} =
\gamma_{db} + {\rm i}(\Delta - \Delta_0 
- \Delta_c)$.

It is instructive to first examine the case of infinitely long lived 
lower level coherence, $\gamma_0 \rightarrow 0$. Fig.2 shows typical 
susceptibility 
spectra in the case of radiative broadening for the system with weak
coherent perturbation of the dark state.  
We note that the original dark resonance is split into a pair of dark lines.
Indeed from Eq.(\ref{lin1}) one finds that the probe susceptibility vanishes 
at the two frequencies:
\begin{eqnarray}
\Delta &=& \Delta_0 + \frac{\Delta_c}{2} \pm \sqrt{\left( \frac{\Delta_c}{2}
\right)^2 + \Omega_c^2}\,,
\end{eqnarray}
i.e., coherent perturbation does not simply eliminate the dark
resonance, but rather splits it into two. These two transparency points
correspond, for resonant coherent fields, to the
dark states of the two $\Lambda$ subsystems of Fig.1b. In addition, a novel 
narrow feature emerges which is superimposed on 
the original transparency line. It is represented by the second term on the 
right-hand side of Eq.~(\ref{lin1}).
In the case of a sufficiently strong driving field 
such that $|\widetilde{\Omega}^2| \gg \Omega_c^2$, this 
feature has an approximately Lorentzian lineshape with line-center
and width given by 
\begin{eqnarray}
\bar{\Delta} \;
\equiv
\; \frac{\Delta_c + \Delta_0}{1 + \Omega_c^2/\widetilde{\Omega}^2},\;
\Gamma_0=\gamma_{ab}\frac{\Omega^2}
{|\widetilde{\Omega}^2|}\frac{2\Omega_c^2}{|\widetilde{\Omega}^2 + 
\gamma_{ab}\Delta_c|},
\end{eqnarray}
where $|\widetilde{\Omega}^2 + \gamma_{ab}\Delta_c| \gg \Omega_c^2$ is assumed.
At $\Delta = \bar{\Delta}$ the susceptibility scales 
like $\chi = \eta/\gamma_{ab}$. 
That means that the amplitude of the  resonances created by the coherent
perturbation is of  the same order as that of
a fully resonant two-level absorber.
The position and width of the absorption line can be
manipulated by tuning ($\Delta_c$) and varying the strength ($\Omega_c$) of
the coherent perturbation. 

The interference nature of the effect becomes most profound 
for $\Omega^2 \approx \Delta_c(\Delta_c + \Delta_0)$, i.e. when the 
energy of one of the dressed
states $\left|+\right\rangle$ or $\left|-\right\rangle$ approaches
the energy of dressed state $\left|0\right\rangle$. In this case the
new feature 
turns from a sharp Lorentzian absorption line into a transparency line
of width
$\sim 2 \Omega_c^2 (\Delta_0+\Delta_c)/(\gamma_{ab}\Delta_c)$ close to the 
point of three-photon resonance (see Fig.2b). 
Hence, by 
proper tuning of the coherent perturbation the multi-photon processes can be
either resonantly enhanced or completely eliminated.   
We emphasize that the ultimate limit for the widths of the described 
high-contrast spectral features in the limit of small Rabi
frequency $\Omega_c$ is determined by the finite relaxation rate of
the long-lived  
coherences between the metastable states. It can therefore be extremely 
small compared to the width of the optical transition. 

We now extend our treatment to include the case when some atoms are 
injected into one of the
levels coupled by the coherent perturbation, specifically into the 
state $\left|d\right\rangle$. In a dressed state picture such
injection implies the  
selective population of the dressed state $\left|0\right\rangle$,
which may result  
in optical gain or in an increase of the refractive index at the 
transparency point.
In the following we focus on the case, where the atoms are 
excited by 
a weak incoherent pump rate $r$. 
Let us consider, for example, the case when all fields are on resonance
(i.e. $\Delta_0 = \Delta_c = \Delta = 0$, see Fig.3a). Assuming,
additionally, a weak 
coherent perturbation and small ground state relaxation 
($\gamma_0\ll |\Omega_c|,r \ll \gamma_{\rm rad}$)  we find that the
absorptive feature 
is turning into gain when 
\begin{eqnarray}
r &\ge& \frac{\gamma_b}{\gamma_d}
\left|\frac{\Omega_c}{\Omega}\right|^2 \gamma_a + 
(1+\frac{\gamma_b}{\gamma_d})\gamma_0  \;.
\end{eqnarray}
Since $|\Omega_c/\Omega|^2$ can be made very small, an incoherent pump
strength orders of magnitude smaller than necessary to invert the
optical transition is sufficient to produce gain.
It is interesting to note that the upper level population in this case 
is very small (${\rho_{aa}^{(0)}} \sim r/\gamma_b$). Furthermore,
double-dark lines can be used to create a  
medium  with an enhanced 
refractive index without absorption. To this end, it is  
favorable to produce a double-dark line at a frequency where the refractive 
index is large in the absence of coherent perturbation, i.e. in the vicinity 
of the dressed states $\left|\pm\right\rangle$.  This can be readily
achieved by tuning the  
coherent perturbation (see Fig.~3b). 
\begin{figure}
\epsfysize=5cm
\center{\leavevmode\epsfbox{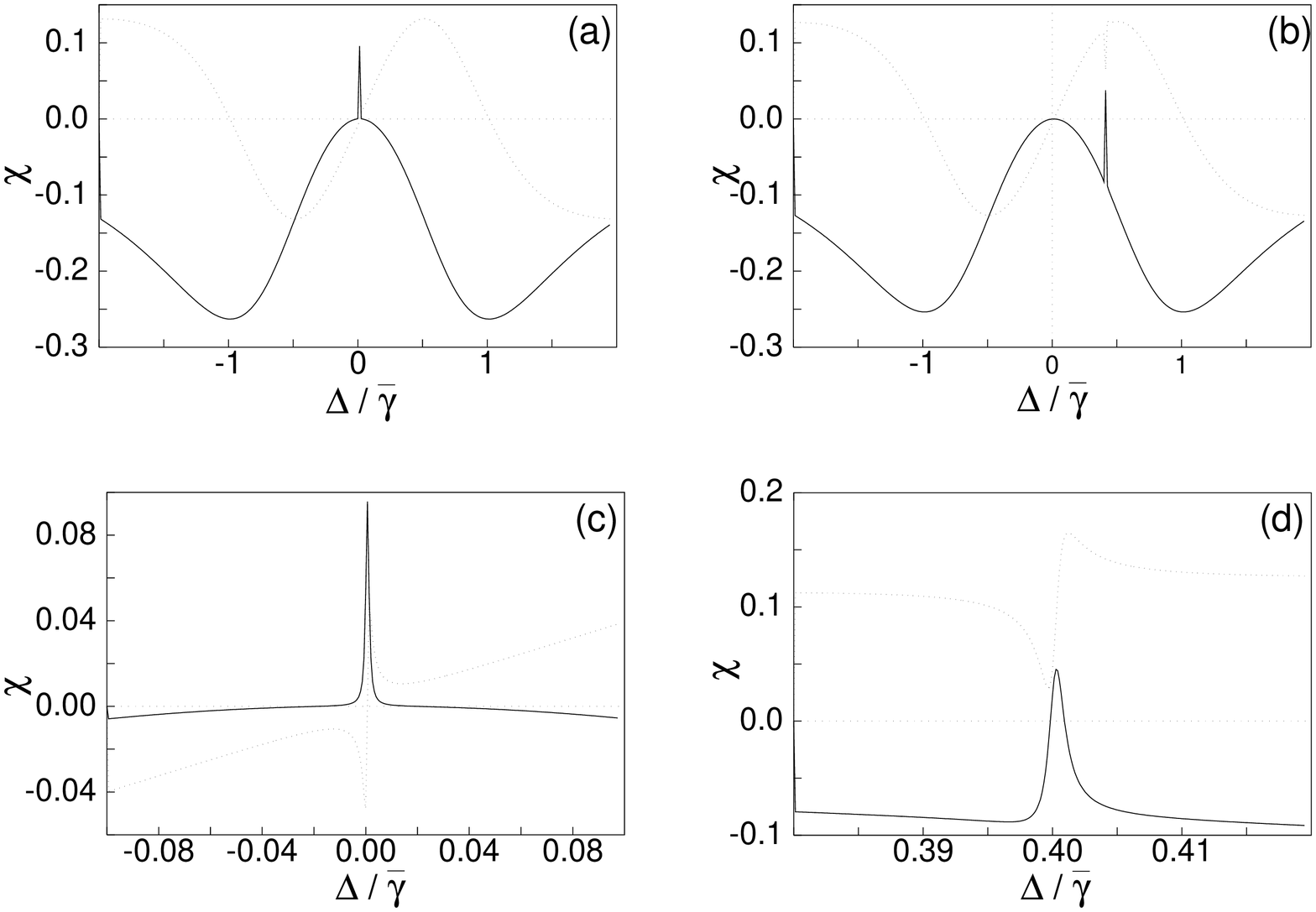}}
\caption{Imaginary (solid lines) and real (dashed lines) part of the
susceptibility in the units of $\eta/\bar{\gamma}$ 
for (a,c) optical gain and (b,d) enhanced index of
refraction without absorption. 
The parameters are $\gamma_b = \gamma_c =
\gamma_d = \bar{\gamma}$, $\Omega = \bar{\gamma}$,
$\Omega_c=0.01\bar{\gamma}$, $\gamma_0 = 3r_b = 
3r_c=3r_d=0.0001\bar{\gamma}$, and $r=0.001\bar{\gamma}$, $\Delta_c=0$
for (a,c) and $r = 
0.0005\bar{\gamma}$, $\Delta_c=0.4\bar{\gamma}$ for
(b,d).}
\end{figure}

Hence, the above considerations allow us to conclude that using coherently 
coupled double dark resonances, it is possible to efficiently ``engineer'' 
the atomic response. We now focus on some of the applications of 
such designed atomic response. Consider, first of all, the problem of optical 
activity enhancement in a dense medium, and in particular, the enhancement of
the refractive index without absorption. One of the major obstacles
in the realization of a large refractive index (i.e. of susceptibility 
$\chi'$ comparable to unity) in usual schemes \cite{hi,indexp} is the
requirement  
of a large excited state population density. 
Excited state population and 
the corresponding  energy dissipation due to spontaneous emission represent an 
important limitation to the optical 
activity enhancement in a dense, partially excited ensemble, as they lead
to the requirement of large continuous energy input, absorption of coherence 
generating fields, superradiant decays, frequency shifts and other 
decoherence effects. 
The potential advantage of using double-dark lines is 
that large susceptibility values can be achieved in dense media 
avoiding the above mentioned problems. We note, for example, that in the 
case depicted in Figs.3(b,d),  the refractive index at the point of
vanishing  
absorption is comparable to the maximum value obtained in a two level 
system in the vicinity of an atomic resonance. It is obtained with a very 
small excitation of atoms and correspondingly small energy
dissipation. The parameters 
used for the present simulations correspond to a possible
realization of double-dark resonances within the Rb $D_1$ absorption line 
using hyperfine and Zeeman sublevels of the ground state. 
Here a pair of optical fields can be used together with an RF or 
a microwave field to 
generate a double dark line. The relaxation rate between metastable lower 
levels in such a system can easily be within few kHz, being limited,
in a dense medium, only by the very slow dephasing due to spin-exchange 
collisions \cite{dop}.
In this case, atomic densities up to $10^{15}-10^{16}$cm$^{-3}$ can be 
used and a few centimeters of transparent Rb vapor with refractivity 
$\chi'\sim 1$  can be created \cite{dop,emph}. This can be put in prospective
by noting that resonant $\chi'\sim 10^{-4}$ was
observed in the experiment of Ref. \cite{indexp}
utilizing a $\Lambda$-type EIT scheme in Rb.

Furthermore, the narrow features
associated with double dark resonances can be of interest in 
high-resolution laser spectroscopy. They can provide a sensitive tool for 
direct measurements of e.g. strength of the coherent perturbation such as 
magnetic fields. High sensitivity can be expected similar to EIT-based
techniques \cite{eitsens} but without the need for involved
dispersive measurements, since  narrow double-dark 
resonances can be observed with a large signal-to-noise
ratio. Indeed, 
high-contrast narrow features persist even in the presence
of strong optical fields and are not limited by power broadening.

Another interesting application of double-dark resonances are unipolar
and bipolar quantum well lasers. Here the properties of double-dark
resonances can provide a way to mitigate the problems associated with
large inhomogeneous broadening \cite{mosconf}.

Finally, the possibility of using double-dark states in adiabatic passage 
 techniques \cite{adi} is intriguing, in that  it offers a way of  
a robust preparation and phase-sensitive probing of arbitrary quantum 
superpositions of lower states \cite{Unanian}, 
which is of particular interest for quantum computation \cite{quco}.
This, as well as other applications of double-dark lines will be addressed in 
detail elsewhere.

The authors gratefully acknowledge useful discussions with 
L.~Hollberg, K.~Bergmann,  E.~Fry, P.~Hemmer, Yu.~Rostovtsev, 
R.~Pfund, V.~Sautenkov, 
B.~W.~Shore, R.~Unanyan, V.~Velichansky and A.~Zibrov and the support from
the Office of Naval Research, the Welch Foundation,
the Texas Advanced Research and Technology Program, the National Science 
Foundation, and the Air Force Laboratories. 
M.~F.~thanks the Alexander-von-Humboldt foundation and S.~F.~Y.~the
``Studienstiftung des Deutschen Volkes'' for financial support.


\begin{references}
\bibitem{emph} We further note that although the atomic response 
of the different systems in Fig.1 on the probe field is similar, 
the propagation dynamics of the other fields can differ. In 
particular, a dense atomic ensemble can give rise to a nonlinear
 generation of new fields components, which can 
significantly modify double-dark lines.
\end{references}
\end{document}